\documentstyle[aps,epsfig,twocolumn]{revtex}

\title{Physical Optimization of Quantum Error Correction Circuits}

\author{
Guido Burkard$^{1*}$,
Daniel Loss$^{1\dagger}$,
David P.\ DiVincenzo$^{2\ddagger}$,
John A.\ Smolin$^{2\S}$}

\address{$^{1}$
Department of Physics and Astronomy,
University of Basel,\\ Klingelbergstrasse 82,
CH-4056 Basel, Switzerland}

\address{$^{2}$
IBM Research Division, T.J.\ Watson Research Center,\\
P.O.\ Box 218, Yorktown Heights, New York 10598}

\newcommand{\xor}{{\sc XOR}}
\newcommand{\cpf}{{\sc CPF}}
\newcommand{\p}{{\cal P}}
\newcommand{\boldphi}{{\mbox{\boldmath $\phi$}}}  
\newcommand{\unity}{{1\!\! 1}}

\begin{document}

\twocolumn[\hsize\textwidth\columnwidth\hsize\csname @twocolumnfalse\endcsname

\maketitle

\begin{abstract}
Quantum error correcting codes have been developed to protect a quantum computer from
decoherence due to a noisy environment.
In this paper, we present two methods for optimizing the physical implementation of such
error correction schemes. First, we discuss an optimal quantum circuit implementation of
the smallest error-correcting code (the three bit code).
Quantum circuits are physically implemented by serial pulses, i.e. by switching on and off
external parameters in the Hamiltonian one after another. In contrast to this, we introduce
a new parallel switching method that allows faster gate operation by switching all external
parameters simultaneously.
These two methods are applied to electron spins in coupled quantum dots subject to a
Heisenberg coupling $H=J(t) {\bf S}_1\cdot {\bf S}_2$ which can generate the universal
quantum gate `square-root-of-swap'.
Using parallel pulses, the encoding for three-bit quantum error correction in a Heisenberg
system can be accelerated by a factor of about two.
We point out that parallel switching has potential applications for arbitrary quantum
computer architectures.
\end{abstract}


\vskip1pc]
\narrowtext


\section{Introduction}\label{introduction}
Quantum computers are capable of efficiently solving problems such as
prime factoring\cite{shor} or simulating other quantum systems\cite{feynman}, for
which no efficient classical algorithm is known.
A quantum computer is a device that stores and processes information which is
physically represented in its quantum state\cite{divincenzo95}.
Typically, such a device contains a collection
of quantum two-state systems, e.g. spin-1/2 particles. The state of each
two-state system then represents a quantum bit, or qubit, the smallest indivisible
unit of information in a quantum computer. Computations are driven by interactions between
the qubits, generating logic gates operating on them. A quantum gate operating on $M$
qubits can be represented as a $2^M\times 2^M$ unitary matrix.
Usually, a computation or algorithm is split up into a series of elementary
gate operations involving only one or two qubits. In this representation, algorithms
are also called quantum circuits. It has been demonstrated that there exist elementary
two-qubit gates $U$ which are universal when complemented with a sufficiently large set
of single-qubit gates\cite{barenco}.
This means that any quantum algorithm can be split up into a quantum circuit 
which contains only $U$ and single-qubit gates. Quantum circuits are in general not the most
efficient way of implementing a quantum computation, as we will demonstrate in this paper.

First experimental realizations of quantum computation using trapped ions\cite{monroe},
optical cavities\cite{turchette}, and NMR\cite{chuang},
involving two or three qubits, have been reported.
In contrary to all of these systems, solid-state implementations have the potential
for a large-scale quantum computer involving hundreds or thousands of qubits.
In this paper, we will concentrate on a theoretical proposal to use coupled semiconductor
quantum dots in which the spin of the excess electron on each dot represents
a qubit\cite{loss}.
Apart from electron spins in quantum dots, a number of other solid-state systems have been
proposed for quantum computation: Nuclear spins of donor atoms in silicon\cite{kane},
Josephson junctions\cite{jj},  d-wave Josephson junctions\cite{ioffe}, and charge degrees
of freedom in quantum dots\cite{charge}.

\begin{figure}
\centerline{\psfig{file=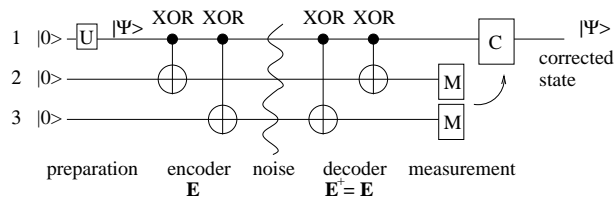,width=8cm}}\bigskip\bigskip
\caption{\label{qec_fig}
The circuit representation for three-bit quantum error correction,
where time is evolving from the left to the right.
First, the three qubits (represented by the horizontal lines) are
initialized. The following unitary transformation $U$ on qubit 1 prepares
the state $|\Psi\rangle$. The encoder $E$ encodes the state $|\Psi\rangle$
in an entangled state of all three qubits. In the next step, 
(simulated) decoherence partly disrupts the state. After the
decoding step (which is identical to the encoding), the qubits 2 and 3
are measured. If they are both one, qubit 1 has to be flipped 
($C=\sigma_x$), otherwise qubit 1 is left unchanged ($C=\unity$).
If no more than one bit flip error occurred, the resulting state
in register 1 is again $|\Psi\rangle$ despite the presence of the
decoherence.}
\end{figure}

The physical implementation of quantum computation hinges upon the ability to find or design
systems in which quantum phase coherence is maintained over long times compared to the
duration of the typical controlled coherent operation. The discovery
of quantum error correcting codes has been a landmark in the effort to find methods to
protect a quantum computer from the decohering effects of a noisy environment\cite{qec}.
The smallest quantum error correcting code for one qubit involving three code qubits
has already been implemented in NMR\cite{cory}.

In this paper, we present theoretical methods for finding an optimal implementation of
three-bit error correction. The optimization is understood here in terms of switching speed
and switching complexity. The former is mainly motivated by the presence of decoherence which makes
fast switching desirable, while the latter can be necessary if the physical implementation
sets limits to the complexity of the switching.
The two optimization goals usually are in conflict to each other, i.e.
a fast implementation usually requires a complex switching mechanism while
switching with a simple mechanism is slow.
First, we will study the ``simple and slow'' switching provided by quantum circuits, and try
to optimize it. Then, we will go on to ``complex and fast'' switching, for which we introduce
parallel (as opposed to serial) pulses for the control parameters of the system, and show that
the parallel pulses allow faster switching than serial pulses. We also introduce a numerical
method for finding such parallel pulses for arbitrary gates and Hamiltonians.
We note that in a similar approach,
Sanders \textit{et al.}\cite{sanders} use a genetic algorithm to find optimized
complex pulses to generate quantum gates using an Ising type Hamiltonian for
optically driven quantum dots coupled by dipole-dipole interactions.
Our approach differs from this work in the underlying Hamiltonian
(i.e. the mechanism proposed for quantum computation); in addition, the parallel
pulses we suggest are general pulses that are discretized in time, whereas
the pulses in \cite{sanders} are chirped Gaussian pulses.

Three-qubit quantum error correction is able to correct either one bit flip or
one phase flip error (correcting both types of errors requires a code with at least
five code qubits\cite{qec}). The operation of the scheme is shown in the quantum
circuit, Fig.~\ref{qec_fig}.
First, all three qubits (represented by horizontal lines) are initialized to the
state $|0\rangle$, e.g. by polarizing the spins with a strong magnetic field.
Then qubit one is rotated into an arbitrary state
$|\Psi\rangle = U|0\rangle = \alpha|0\rangle + \beta|1\rangle$.
The purpose of the quantum error correction scheme is to protect this state from
external decoherence.
In order to do this, the state $|\Psi\rangle$ is encoded in the highly entangled
three-qubit state $|\Psi_L\rangle = E(|\Psi\rangle_1|0\rangle_2|0\rangle_3)
= \alpha|0\rangle_1|0\rangle_2|0\rangle_3 + \beta|1\rangle_1|1\rangle_2|1\rangle_3$.

The encoded state $|\Psi_L\rangle$ can then be subject to external noise causing
a (partial) bit flip of one of the spins, $|\Psi_L\rangle\rightarrow
\exp(i\epsilon \sigma^x_i)|\Psi_L\rangle=|\Psi_L'\rangle$ without the information
contained in the encoded state $|\Psi_L\rangle$ being lost.
The state $|\Psi_L\rangle$ is recovered (decoded) by applying the inverted encoding
network $E^{-1}=E^\dagger=E$ which is identical to $E$ since $E$ is Hermitian.
Then the qubits 2 and 3 are measured (for physical implementations of quantum
measurements on spins in quantum dots,
see Refs.~\onlinecite{loss,burkard,divincenzo98b,divincenzo99}.
If both qubits are in state $|1\rangle$, then
qubit 1 is flipped, otherwise it is left unchanged. This restores the
state $|\Psi\rangle$ in qubit 1 which then can be measured in order to 
check the functionality of the scheme.

For a first experiment one would probably want sufficiently low noise such that
the state $|\Psi\rangle$ is not destroyed by ``natural'' noise. One would then
introduce bit flips ``by hand'' by applying a random oscillatory magnetic
field in the $x$ direction and then check whether those artificial errors
can be corrected for.

For the sake of concreteness, we will apply our methods to a system of coupled spins ${\bf S}_i$
with $s=1/2$ (each representing a qubit), subject to isotropic spin-spin interaction and local
magnetic fields. With this model we capture the physics of electrons in coupled quantum
dots\cite{loss}. We emphasize that a generalization to other systems with different Hamiltonians
is straight-forward and does not require a new method for optimizing the switching process.

Our paper is organized as follows:
In Sec.~\ref{model}, we introduce the formalism that we use to describe the 
dynamics of electron spins in coupled quantum dots and other Heisenberg systems.
The methods developed in Secs.~\ref{serial_pulse} and \ref{parallel_pulse},
including the use of parallel pulses, are not special to the Heisenberg 
Hamiltonian Eq.~(\ref{hamiltonian}). As an example we give some results
for transversely coupled spins (XY model) in Section \ref{anisotropic},
because they are encountered when electronic spins are coupled using
cavity-QED\cite{imamoglu}.
The results of the Sections \ref{serial_pulse}-\ref{anisotropic} are independent
of the mechanisms that are involved in their physical implementation -  they
are derived under the assumption that the model Hamiltonian Eq.~(\ref{hamiltonian})
(or Eq.~(\ref{xy_hamiltonian})) is exact. In Section \ref{requirements} we
discuss some limitations and necessary conditions for the validity of
this approach. Finally, in Sec.~\ref{applications}, we give a detailed list
of instructions for both serial and parallel switching which must be followed
in order to implement three-qubit quantum error correction in a system of
spins subject to Heisenberg interactions in experiment.

\section{Model}\label{model}
In the system we consider the qubit is represented by the spin 1/2 state of the excess electron
in a quantum dot, i.e. the ``spin up'' state $|\!\!\uparrow\rangle$ is identified with the
logic state $|0\rangle\equiv|\!\!\uparrow\rangle$ and likewise
$|1\rangle\equiv|\!\!\downarrow\rangle$,
where the quantization axis is chosen along the $z$ axis,
$\sigma^z|\!\!\uparrow\rangle=+|\!\!\uparrow\rangle$ and
$\sigma^z|\!\!\downarrow\rangle=-|\!\!\downarrow\rangle$.

The excess electron spins in a pair of quantum dots which are linked through
a tunnel junction can be described by the Heisenberg Hamiltonian\cite{loss,burkard}
\begin{equation}\label{hamiltonian}
H(J,{\bf B}_i,{\bf B}_j) = J\,\,{\bf S}_i\cdot{\bf S}_j
+ {\bf B}_i\cdot{\bf S}_i+ {\bf B}_j\cdot{\bf S}_j,
\end{equation}
where ${\bf S}_i={\bf \sigma}_i/2$ describes the (excess) spin 1/2 on dot $i$ and $J$ denotes the
exchange energy, i.e. the
energy gap between the spin singlet and triplet states\cite{loss}. This effective Hamiltonian can
be derived from a microscopic model for electrons in coupled quantum dots\cite{burkard},
see also Sec.~\ref{requirements}. It is found that
$J$ can be changed using a variety of external parameters. Tuning the gate voltage between the coupled
dots changes the height of the tunneling barrier and therefore directly alters $J$. Note that
$J$ is exponentially sensitive to barrier changes. Also, 
applying a magnetic field perpendicular to the 2DEG within which the quantum dots are defined greatly 
influences the exchange coupling $J$ and can even result in a sign change of $J$ for unscreened
Coulomb interaction\cite{burkard}.
Some coupling of the spin ${\bf S}_i$ to a local external magnetic field ${\bf B}_i$ is also necessary
for quantum computation, and
has been included in the Hamiltonian Eq.~(\ref{hamiltonian}).
Note that we have included the factor $g_i\mu_B$ in the definition of the magnetic field ${\bf B}_i$,
where $g_i$ is the g-factor for dot $i$ and $\mu_B$ is the Bohr magneton.
The physical realization of the field
gradients or inhomogeneous g-factors required for the local magnetic fields is challenging, but
there exist several possibilities for generating them \cite{burkard,divincenzo99}.

\begin{figure}
\centerline{\psfig{file=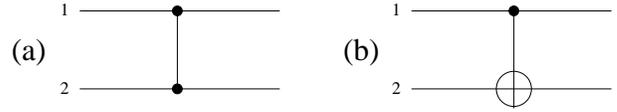,width=8cm}}\bigskip\bigskip
\caption{\label{ugates}
Circuit notation of two universal gates:
(a) The `square-root-of-swap' (S) gate, (b) the XOR gate.}
\end{figure}

From the Hamiltonian Eq.~(\ref{hamiltonian}) we can generate the following set of quantum gates,
\begin{eqnarray}
U_i(\boldphi) &=& \exp(i \boldphi \cdot{\bf S}_i),\label{single_qubit}\\
S &\equiv& S(i,j) \equiv U_{\rm swap}^{1/2} = e^{-i\pi/8}\exp\left(i\frac{\pi}{2}{\bf S}_i\cdot{\bf S}_j\right)\label{sqrt_swap}.
\end{eqnarray}
The single-qubit operation $U_i(\boldphi)$ for the spin ${\bf S}_i$ is generated by applying a magnetic field pulse ${\bf B}(t)$ at the location of the spin ${\bf S}_i$ such that $\int_0^t d\tau {\bf B}(\tau)=\boldphi$. Similarly, the `square-root of swap' gate\cite{loss} (which we denote by $S$ in the following) is obtained by switching the interaction $J(t)$ between the spins ${\bf S}_i$ and ${\bf S}_j$ such that $\int_0^t d\tau J(\tau)=\pi/2$. We introduce the circuit notation for $S$ in Fig.~\ref{ugates}a. Note that $U_i(\boldphi)$ is $4\pi$-periodic in $\phi$, and $2\pi$-periodic up to a global phase $-1$, which for our purposes is not important. Eqs.~(\ref{single_qubit}) and (\ref{sqrt_swap}) are a universal set of gates. Other powers $U_{\rm swap}^\alpha$ of the swap gate $U_{\rm swap}=S^2=e^{-i\pi/4}\exp(i\pi{\bf S}_i\cdot{\bf S}_j)$: ($|ab\rangle = |a\rangle_i\otimes|b\rangle_j$)
\begin{equation}\label{swap}
|00\rangle\mapsto|00\rangle,\;|01\rangle\mapsto|10\rangle,\;
|10\rangle\mapsto|01\rangle, \; |11\rangle\mapsto|11\rangle,
\end{equation}
can also be generated by the Hamiltonian Eq.~(\ref{hamiltonian}), but are not necessary for universality, once $S=U_{\rm swap}^{1/2}$ is included. Note that we cannot substitute $U_{\rm swap}^{1/2}$ by any other power of $U_{\rm swap}$ because only $U_{\rm swap}^{\alpha}$ with $\alpha$ half an odd integer has the ability to produce maximal entanglement from an unentangled state.

\begin{figure}
\centerline{\psfig{file=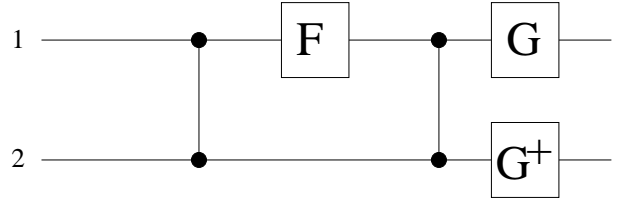,width=8cm}}\bigskip\bigskip
\caption{\label{cpf_fig}
A circuit representation for the conditional phase flip ($\cpf$), Eq.~(\ref{CPF}),
as given in Eq.~(\ref{cpf_circuit}). The single qubit rotations are $F=e^{i\pi S^z}$
and $G=e^{i\frac{\pi}{2}S^z}$. The $\cpf$ is related to the XOR gate Eq.~(\ref{XOR})
by the basis transformation Eq.~(\ref{basis_change}).}
\end{figure}

\begin{figure}
\centerline{\psfig{file=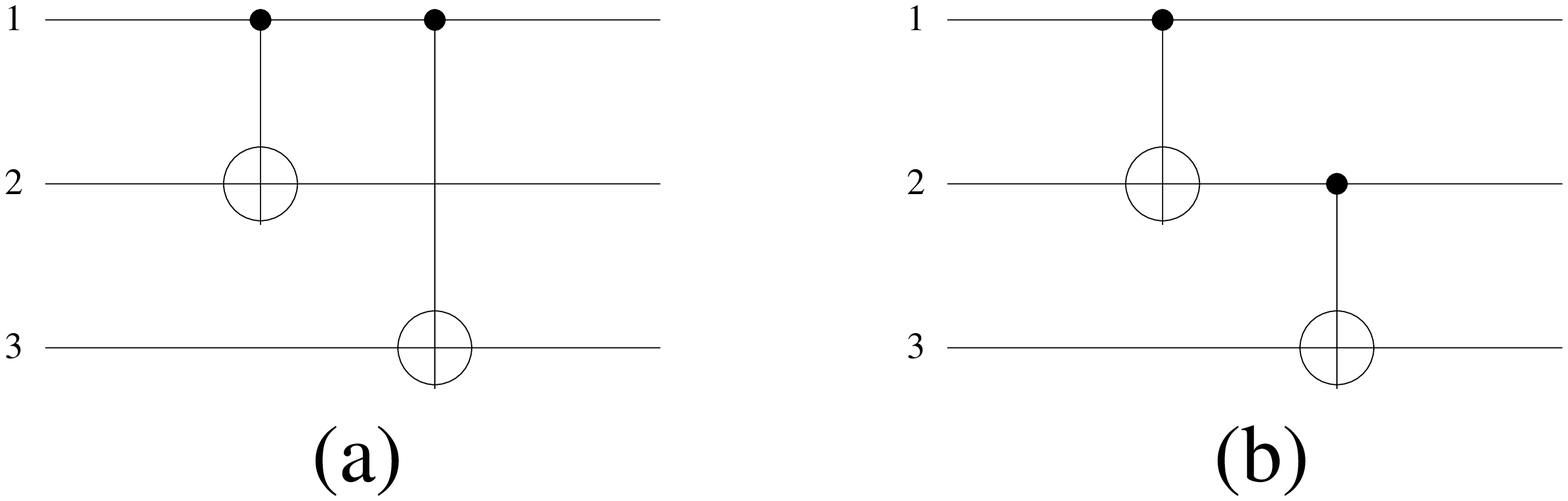,width=8cm}}\bigskip\bigskip
\caption{\label{encoder}
The quantum circuits for (a) the three-bit encoder $E$, cf.\ Eq.~(\ref{e_circuit}),
and (b) the teleportation encoder $E_T$, cf.\ Eq.~(\ref{et_circuit}).}
\end{figure}

We can use `square-root of swap' to generate the controlled phase flip gate $U_{\cpf}$:
\begin{equation}\label{CPF}
|00\rangle\mapsto|00\rangle,\,|01\rangle\mapsto|01\rangle,
\,|10\rangle\mapsto|10\rangle,\,|11\rangle\mapsto-|11\rangle,
\end{equation}
with the quantum circuit depicted in Fig.~\ref{cpf_fig} (time evolving from the left to the right),
or formally\cite{loss},
\begin{equation}\label{cpf_circuit}
U_{\cpf}=e^{-i\frac{\pi}{2}}e^{i\frac{\pi}{2}S_1^z}e^{-i\frac{\pi}{2}S_2^z} S e^{i\pi S_1^z} S,
\end{equation}
which in turn is related to $\xor$ gate $U_{\xor}$:(Fig.~\ref{ugates}b)
\begin{equation}\label{XOR}
|00\rangle\mapsto|00\rangle,\;|01\rangle\mapsto|01\rangle,
\;|10\rangle\mapsto|11\rangle, \; |11\rangle\mapsto|10\rangle,
\end{equation}
by the basis change
\begin{eqnarray}
U_{\xor} &=& V  U_{\cpf}  V^{\dagger},\label{basis_change}\\
V       &=& \exp(-i\pi S_2^y/2).\nonumber
\end{eqnarray}
Since $\xor$ with one-bit gates is a universal quantum gate\cite{barenco}, this confirms that
Eqs.~(\ref{single_qubit}) and (\ref{sqrt_swap}) are a universal set of gates.
In what follows, we will use the $\xor$ gate to construct the gate $E$ that performs the
encoding for three-qubit quantum error correction, as shown in Fig.~\ref{qec_fig}, and
can be obtained by two successive $\xor$ gates (Fig.~\ref{encoder}a),
\begin{equation}\label{e_circuit}
E=U_{\xor}(1,3)   U_{\xor}(1,2).
\end{equation}
The very similar quantum gate (Fig.~\ref{encoder}b)
\begin{equation}\label{et_circuit}
E_T=U_{\xor}(2,3)   U_{\xor}(1,2),
\end{equation}
has the nice property that it can be used for
implementing the quantum teleportation of one qubit as a quantum
computation\cite{brassard}. It is clear that our analysis of the $\xor$
gate can also be used for implementing this gate.


\section{Serial pulse mode}\label{serial_pulse}

In the foregoing discussion we made clear why it is desirable to generate certain quantum gates or networks such as $\xor$, $E$, and $E_T$, and that it is indeed possible to produce them using a system of spins that are mutually coupled by the Heisenberg interaction Eq.~(\ref{hamiltonian}). In fact, we know that we can generate every quantum gate using those interactions, since Eqs.~(\ref{cpf_circuit}) and (\ref{basis_change}) explicitly tell us how to produce $\xor$, which together with the set of single-qubit operations forms a universal set of gates for quantum logic\cite{divincenzo98a}. We now go one step further and ask ourselves which is the {\em most efficient} way of implementing a certain quantum gate. More precisely, we are interested in minimizing the switching time $\tau_s$ for the desired quantum gate. This kind of optimization is crucial because the error probability per gate operation is proportional to the switching time, $\epsilon=\tau_s/\tau_\phi$, where $\tau_\phi$ denotes the dephasing time of the system. Other criteria for optimization can be added, if e.g. one kind of elementary task (say, spin-spin interactions) is much harder to perform than another (such as single-qubit rotations), or if the switching of parameters turns out to be difficult.

For the purpose of finding an optimal implementation of quantum gates, we first define which set of elementary operations we are going to use. We will call this set the serial pulse operations, since they can be achieved by `switching on' exactly one of the parameters $\vec{p}=(J, {\bf B}_1, {\bf B}_2)$ in the Hamiltonian Eq.~(\ref{hamiltonian}) for some finite time. Clearly, $\xor$ does not belong to this class of gate operations -- it takes the whole sequence Eq.~(\ref{cpf_circuit}) to produce it. Note that the definition of serial (and later parallel) pulse operations depends on the Hamiltonian and on how we parametrize it. The distinction only makes sense if serial pulse operations correspond to physically switching on and off a part of the device, e.g. the magnetic field at the location of one of the spins. We will use the serial pulse operations defined in Eqs.~(\ref{single_qubit}) and (\ref{sqrt_swap}).

\subsection{$\xor$ gate}
As a first example for our efficiency analysis we take the sequence Eq.~(\ref{cpf_circuit}) for $\xor$. We do not try to optimize the length of the single-qubit pulses. Instead we investigate whether it really takes two instances of $S$ or if $\xor$ can be performed with one $S$ plus single-qubit operations. This is most reasonable if coupling two qubits is more costly (e.g. due to decoherence) than operating on a single qubit. Thus, our question is whether
\begin{equation}\label{xor_trial}
S_1 = \left[u_{21}\otimes u_{22}\right]  S  \left[u_{11}\otimes u_{12}\right],
\end{equation}
is equal to $U_{\xor}$ for some choice of single-qubit gates $u_{nm}=U_m(\boldphi_{nm})$ or not.
We will shortly prove that the answer is negative and it indeed takes at least two $S$ to produce $\xor$, but first we introduce the method we developed in order to prove this kind of `no-go' theorem. The idea is that very often quantum gates can be distinguished by their ability to produce entanglement. This property of quantum gates has the advantage that it is invariant under concatenation with arbitrary single-qubit gates.

We denote the product (pure) states in our two-qubit Hilbert space ${\cal H}={\cal H}_2^{\otimes 2}$ by $\p = \left\{|\Psi\rangle\in{\cal H}\Big| \, |\Psi\rangle=|\varphi\rangle\otimes|\chi\rangle; \,|\varphi\rangle, |\chi\rangle\in{\cal H}_2\right\}$\cite{footnote0}. Here, ${\cal H}_2$ denotes the single-qubit Hilbert space with basis $|0\rangle$, $|1\rangle$. A state $|\Psi\rangle\notin \p$ is called entangled. For every quantum gate (unitary matrix) $U$ acting on ${\cal H}$, we define the subset $\p(U)=\left\{|\Psi\rangle\in\p\Big|\, U|\Psi\rangle\in\p\right\}\subseteq \p$ of product states which are mapped onto a product state by $U$. The idea is now simply that two quantum gates $U_1$ and $U_2$ which have different $\p$-sets, $\p(U_1)\neq\p(U_2)$, obviously must be different: $U_1\neq U_2$ (note that this implication cannot be reversed). The $\p$-set of $\xor$ is
\begin{equation}
\p(U_{\xor})=\left\{|0\phi\rangle, |1\phi\rangle, |\phi\pm\rangle \Big| |\phi\rangle\in{\cal H}_2\right\},
\end{equation}
where we used the notation $|\pm\rangle=(|0\rangle\pm|1\rangle)/\sqrt{2}$.

In order to find $\p(S)$ it is useful to convince oneself that `square-root of swap' operates on a product state $|\phi\chi\rangle$ according to the very intuitive formula
\begin{equation}\label{s_operation}
S|\phi\chi\rangle = \frac{1}{1+i}\left(|\phi\chi\rangle + i|\chi\phi\rangle\right),
\end{equation}
by first checking it for the basis of products of $|0\rangle$ and $|1\rangle$, and then using that the right-hand side of Eq.~(\ref{s_operation}) is linear in $|\phi\rangle$ and $|\chi\rangle$.
From this rule we conclude that all product states become entangled by $S$ unless they are the product of two equal single-qubit states,
\begin{equation}\label{s_pset}
\p(S)=\left\{|\Psi\rangle\in\p \Big| \exists|\phi\rangle\in{\cal H}_2 : \,|\Psi\rangle=|\phi\phi\rangle\right\}.
\end{equation}
For any choice of the $u_{nm}$ in Eq.~(\ref{xor_trial}), we can construct the state $|0\rangle\otimes u_{12}^\dagger u_{11} |1\rangle$ which is in $\p(U_{\xor})$ but not in $\p(S_1)$ since $S_1|\Psi\rangle$ is entangled. Therefore, $\p(S_1)\neq\p(U_{\xor})$ and consequently $S_1\neq U_{\xor}$ for any choice of $u_{nm}$. Thus, the sequence given in Eq.~(\ref{cpf_circuit}) is optimal in the sense that both `square-root of swap' operations are really needed.
Allowing arbitrary powers of $U_{\rm swap}$ does not reduce the number of two-qubit gates required for $\xor$ either, since one $U_{\rm swap}^{\alpha}$, where $\alpha$ is not an even multiple of $1/2$, cannot act as a perfect entangler which is required for the $\xor$ gate. For completeness, we give here the generalization of Eqs.~(\ref{s_operation}) and (\ref{s_pset}) for arbitrary powers of swap,
\begin{eqnarray}
U_{\rm swap}^{\alpha}|\phi\chi\rangle &=& e^{-i\pi\alpha/2}\left(\cos\left(\frac{\pi\alpha}{2}\right)|\phi\chi\rangle+i\sin\left(\frac{\pi\alpha}{2}\right)|\chi\phi\rangle\right),\nonumber \\
\p(U_{\rm swap}^{\alpha}) &=& \left\{\begin{array}{l l}\p(S),&\alpha\neq\mbox{integer},\\ \p,&\alpha =\mbox{integer}.\end{array}\right.
\end{eqnarray}

\subsection{Three bit encoder $E$}
Regarding the three bit encoder $E$, Eq.~(\ref{e_circuit}), our result tells us that the straight-forward implementation of $E$ requires `square-root of swap' four times, i.e. twice for every $\xor$. This does not mean that there cannot be a more efficient implementation of $E$ than given in Eq.~(\ref{e_circuit}). We can try to implement $E$ using the serial pulse gate set instead of $\xor$'s. It turns out that this is impossible with fewer than four $S$ gates. The analysis still relies on the previously introduced $\p$-set but is slightly more complicated than the one for $\xor$ since in the case of three qubits each gate $S$ can be applied to one of three possible pairs of qubits.

It is clear that just one use of $S$ (plus arbitrary single-qubit operations) cannot produce $E$,
\begin{equation}\label{s1_trial}
U_1 = [u_{21}\otimes u_{22}\otimes u_{23}]  S(i,j) [u_{11}\otimes u_{12}\otimes u_{13}]
\neq E,
\end{equation}
for any choice of $u_{nm}$ for the simple reason that $E$ is able to entangle the qubit 1 with 2, and also 1 with 3, whereas $S(i,j)$ can only entangle the qubits $i$ and $j$ with each other (at most one pair).

It is less obvious that none of the sequences
\begin{equation}\label{s2_trial}
U_2 = U^{(3)}  S(k,l)  U^{(2)}  S(i,j)  U^{(1)},
\end{equation}
with $U^{(n)}=u_{n1}\otimes u_{n2}\otimes u_{n3}$ can reproduce $E$. The idea of the following argument is the same as for the one for $\xor$: We are seeking a state $|\Psi\rangle$ that becomes entangled when acted on with the operator $U_2$ given in Eq.~(\ref{s2_trial}) but remains unentangled under the operation $E$, i.e. $|\Psi\rangle\in\p(E)\setminus\p(U_2)$. This is the case if $|\Psi\rangle$ is entangled by $S(i,j)$ and not disentangled by $S(k,l)$.
We can exclude the case where $(k,l)=(i,j)$ or $(k,l)=(j,i)$, using the same argument as for $U_1$, defined in Eq.~(\ref{s1_trial}).
In the remaining cases it is clear that if $|\Psi\rangle\in\p(E)$ is chosen such that it is entangled by $S(i,j)$, then it will not be disentangled again by $S(k,l)$, and we are done.
Since Eq.~(\ref{s2_trial}) is invariant when $i$ and $j$ (or $k$ and $l$) are interchanged, we can always arrange that $i$, $j$, and $k$ are mutually different, and $l=i$. In the case where $j\neq 1$ the state $|\Psi\rangle = |0\rangle_i\otimes u_{1j}^\dagger u_{1i}|1\rangle_j\otimes |0\rangle_k\in\p(E)$ is not in $\p(U_2)$, because the entanglement between qubit $j$ and the qubits $i$ and $k$ created by $S(i,j)$ cannot be undone by $S(i,k)$. For $j=1$ we choose the state $|\Psi\rangle = u_{1i}^\dagger u_{1j}|1\rangle_i\otimes |0\rangle_j\otimes |0\rangle_k\in\p(E)$ with the same property. This concludes our proof that there is no circuit $U_2$ of the form Eq.~(\ref{s2_trial}) which is equal to $E$. Note that this conclusion is independent of the choice of single-qubit operations in $U_2$, hence the inequality we proved concerns all circuits of the type $U_2$.

\begin{figure}
\centerline{\psfig{file=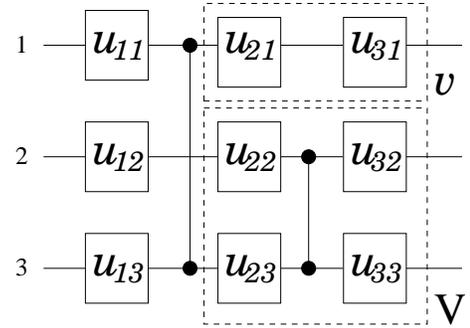,width=6cm}}\bigskip\bigskip
\caption{\label{twoS}
Quantum circuits of the type described in Eq.~(\ref{s2_example}).
Dashed lines represent the grouping in Eq.~(\ref{s2_regrouped}).
}
\end{figure}

As an example consider the circuit
\begin{equation}\label{s2_example}
U_2 = U^{(3)}  S(2,3)  U^{(2)}  S(1,3)  U^{(1)},\\
\end{equation}
for which $i=3$, $j=1$, and $k=2$.
By rewriting $U_2$ in the form shown in Fig.~\ref{twoS},
\begin{equation}\label{s2_regrouped}
U_2 = \left[ v\otimes V(2,3)\right]  S(1,3) U^{(1)},
\end{equation}
with $v=u_{31}u_{21}$ and $V(2,3)=(\unity\otimes u_{32}\otimes u_{33})S(2,3)(\unity\otimes u_{22}\otimes u_{23})$,
and using Eq.~(\ref{s_operation}),
we observe that the product state $|\Psi\rangle =|0\rangle_1\otimes |0\rangle_2\otimes u_{13}^\dagger u_{11}|1\rangle_3\in \p(E)$ is mapped to
\begin{equation}\label{s2_output}
U_2|\Psi\rangle = \frac{1}{1+i} \Big(v|\alpha\rangle_1\otimes V|\delta\rangle_{23} +iv|\beta\rangle_1\otimes V|\gamma\rangle_{23} \Big).
\end{equation}
The unitarity of $u_{11}$ implies that $|\alpha\rangle =u_{11}|0\rangle$ is orthogonal to $|\beta\rangle=u_{11}|1\rangle$, and $|\gamma\rangle = u_{12}|0\rangle\otimes|\alpha\rangle$ is orthogonal to $|\delta\rangle = u_{12}|0\rangle\otimes|\beta\rangle$.  The gates $v$ and $V$ are also unitary, thus $v|\alpha\rangle$ is orthogonal to $v|\beta\rangle$, and $V|\gamma\rangle$ is orthogonal to $V|\delta\rangle$, which implies that $U_2|\Psi\rangle$, Eq.~(\ref{s2_output}), is an entangled state between subsystem 1 and subsystem 2 and 3. From this we conclude that $|\Psi \rangle\in\p (E)$ is not in $\p (U_2)$, and therefore $E\neq U_2$.

\begin{figure}
\centerline{\psfig{file=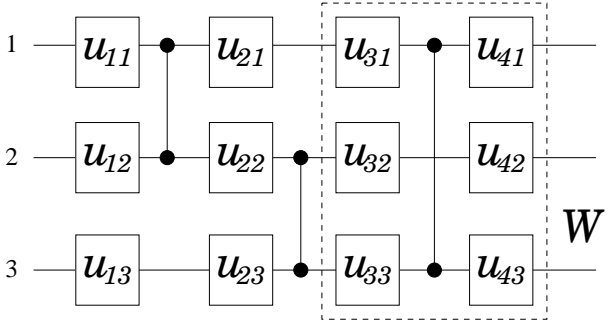,width=8cm}}\bigskip\bigskip
\caption{\label{threeS}
Quantum circuits of the type described in Eq.~(\ref{e_trial_3_example}). Dashed lines represent the grouping in Eq.~(\ref{e_trial_3_regrouped}).
}
\end{figure}

Next, we develop a proof that even with the use of three $S$ gates, $E$ cannot be implemented.
Since each $S$ can couple one of three possible pairs $i_k,j_k=1,...,3$, $i_k\neq j_k$, of qubits,
there are $3^3=27$ sequences including three `square-root of swap' ($S$) gates, having the form
\begin{equation}\label{s3_trial}
U_3 = U^{(4)} S(i_3,j_3) U^{(3)} S(i_2,j_2) U^{(2)} S(i_1,j_1) U^{(1)},
\end{equation}
with arbitrary single-qubit gates $U^{(n)}=u_{n1}\otimes u_{n2}\otimes u_{n3}$.
First we observe that if $(i_2,j_2)=(i_3,j_3)$ or $(i_2,j_2)=(j_3,i_3)$ then we can apply the same argument as for circuits of the type $U_2$ with $(i,j)=(i_1,j_1)$ and $(k,l)=(i_2,j_3)$. In the case where the first two $S$ gates (but not the third one) act on the same pair of qubits, $(i_1,j_1)=(i_2,j_2)$ or $(i_1,j_1)=(j_2,i_2)$, we note that either $|0\rangle_{i_1}|0\rangle_{j_1}|0\rangle_{k}\in\p(E)$ or $|0\rangle_{i_1}|0\rangle_{j_1}|1\rangle_{k}\in\p(E)$, where $k\neq i_1, j_1$, becomes entangled by $U_3$. Therefore $U_3\neq E$ if the first two or the last two $S$ gates operate on the same pair of qubits. In all other cases, we can label the three qubits with three distinct numbers $a$, $b$, and $c$ between 1 and 3 such that $S(i_1,j_1)=S(a,b)$, $S(i_2,j_2)=S(a,c)$, and $S(i_3,j_3)=S(b,x)$, with $x=a$ or $x=c$. The state $|\Psi\rangle$, defined as 
\begin{equation}\label{s3_proof}
\begin{array}{l l}
|0\rangle_{a} \otimes u_{1 b}^\dagger u_{1a}|0\rangle_{b} \otimes u_{1c}^\dagger u_{2c}^\dagger u_{2a}u_{1a}|1\rangle_{c}, & ($\rm if$\; a=1),\\
u_{1a}^\dagger u_{1b}|0\rangle_a \otimes |0\rangle_b \otimes u_{1c}^\dagger u_{2c}^\dagger u_{2a}u_{1b}|1\rangle_{c}, & ($\rm if$\; b=1),\\
u_{1a}^\dagger u_{2a}^\dagger u_{2c} u_{1c} |0\rangle_a \otimes u_{1b}^\dagger u_{2a}^\dagger u_{2c} u_{1c} |0\rangle _b \otimes |1\rangle _c, & ($\rm if$\; c=1),\\
\end{array}
\end{equation}
is chosen such that $|\Psi\rangle \in \p(E)$ and has the property that $S(a,b) U^{(1)}|\Psi\rangle$ is unentangled, but in $S(a,c) U^{(2)} S(a,b) U^{(1)}|\Psi\rangle$ there is entanglement between the qubits  $a$ and $c$. Finally, the $S(b,x)$ gate cannot remove the entanglement; in the final state $U_3|\Psi\rangle$, either the qubit $a$ (if $x=c$) or the qubit $c$ (if $x=a$) is entangled with the other two qubits, thus $|\Psi\rangle\notin\p(U_3)$. Since $|\Psi\rangle\in\p(E)$, this concludes our proof of the statement $U_3\neq E$. In order to illustrate our proof, we apply it to the specific example ($a=2$, $b=1$, $c=3$, $x=c=3$)
\begin{equation}\label{e_trial_3_example}
U_3 = U^{(4)}  S(1,3)   U^{(3)}  S(2,3)   U^{(2)}  S(1,2)  U^{(1)},
\end{equation}
which can be written in the form (shown in Fig.~\ref{threeS})
\begin{equation}\label{e_trial_3_regrouped}
W S(2,3) U^{(2)} S(1,2) U^{(1)},
\end{equation}
where $W$ is a gate that does not couple qubit 2 with any other qubit.
Applying this operator to the state $|\Psi\rangle = |0\rangle_1\otimes
u_{12}^\dagger u_{11}|0\rangle_2\otimes u_{13}^\dagger u_{23}^\dagger u_{22} u_{11}|1\rangle_3$
proves that $U_3\neq E$ for any choice of the $U^{(n)}$, because $U_3|\Psi\rangle$ is entangled,
$|\Psi\rangle\notin \p(U_3)$.

\subsection{Teleportation encoder $E_T$}
Now we consider the teleportation encoder gate $E_T$, Eq.~(\ref{et_circuit}), which is shown in Fig.~\ref{encoder}b. The gate $E_T$ consists of two $\xor$ gates, but in contrast to $E$ these two $\xor$s are arranged in a less symmetric way.
As a consequence, $\p(E_T)$ is only a subset of (and not equal to) $\p(E)$, since e.g. $|0+0\rangle$ is in $\p(E)$ but not in $\p(E_T)$. Again, we can ask whether it is feasible to assemble $E_T$ using less than the four $S$ gates that are used when we simply combine Eqs.~(\ref{cpf_circuit}), (\ref{basis_change}), and (\ref{et_circuit}). The answer is again negative. The proof of this statement is similar to the one given for $E$. It is again clear that a circuit involving one $S$ cannot entangle any pair of the three qubits, as it should in order to reproduce $E_T$.

A circuit of the form $U_2$, Eq.~(\ref{s2_trial}), involving two $S$ gates cannot be equal
to $E_T$ either. We first note that if $S(i,j)=S(k,l)$, then $U_2$ cannot produce entanglement between the third qubit $m\neq i,j$ and $i$ or $j$, and thus $U_2\neq E_T$, because $E_T$ can entangle any pair of qubits.
If $S(i,j)\neq S(k,l)$, we find that either $|\Psi\rangle =|0\rangle_i|0\rangle_j|0\rangle_k\in\p(E_T)$ or $|\Psi\rangle =|1\rangle_i|0\rangle_j|0\rangle_k\in\p(E_T)$, where $k\neq i,j$, is entangled by the gate $S(i,j)$. Since this entanglement cannot be undone by $S(k,l)\neq S(i,j)$, the state $U_2|\Psi\rangle$ is entangled. Thus, we have found a state which is in $\p(E_T)$ and not in $\p(U_2)$ and therefore $U_2\neq E_T$.

We finally explore whether a circuit $U_3$ containing three $S$ gates as in Eq.~(\ref{s3_trial}) can reproduce $E_T$. For $S(i_2,j_2)=S(i_3,j_3)$ we can see that this is not the case by applying the same argument as above for a circuit with two $S$. In the opposite case, $S(i_2,j_2)\neq S(i_3,j_3)$, either the state $|0\rangle_{i_1}|0\rangle_{j_1}|0\rangle_k\in\p(E_T)$ or the state $|0\rangle_{i_1}|0\rangle_{j_1}|1\rangle_k\in\p(E_T)$, with $k\neq i_1,j_1$, is entangled by $S(i_2,j_2)$ or $S(i_3,j_3)$. Because $S(i_3,j_3)\neq S(i_2,j_2)$, entanglement produced by $S(i_2,j_2)$ is not undone by $S(i_3,j_3)$ and therefore we have found a state in $\p(E_T)$ which is not in $\p(U_3)$, completing the proof for $U_3\neq E_T$ for the case of three $S$ gates. This finally implies that, like $E$, the teleportation encoder $E_T$ cannot be constructed using fewer than four $S$ gates.

\subsection{Numerical search}
The method that was presented for proving inequalities between two gates $U_1$ and $U_2$ involving the sets $\p(U_1)$ and $\p(U_2)$ has the advantage that it yields rigorous results although we do not know the details about the involved single-qubit operations. Sometimes however, proofs become lengthy and rather unsystematic, so we would like to have a better tool for the complex cases. Unfortunately, we do not have such a tool which is capable of giving rigorous proofs for inequalities, like the $\p$-set method. However, we have developed a computer algorithm that searches for the $M$ qubit gate $U_g$ in the set
\begin{equation}\label{minimizer_trial}
U\{U^{(n)}\} = U^{(N+1)} X_N U^{(N)} \cdots X_2 U^{(2)} X_1 U^{(1)},
\end{equation} 
where the $X_n$ are arbitrary but fixed $2,3,...,M$ qubit gates and $U^{(n)}=u_{n1}\otimes\cdots\otimes u_{nM}$ are arbitrary and variable products of single-qubit gates (several $u_{nk}$ can be unity). A result of a numerical search is a list $\{ u_{nk}\}$ of single-qubit gates, which satisfy the equation $U\{U^{(n)}\}=U_g$. The computer algorithm can therefore (in the case of a successful run) `prove' equalities, but one or several unsuccessful runs do not constitute a proof that the gate $U_g$ cannot be constructed with a given sequence $X_n$, $n=1,...,N$. Note that the situation is thus exactly opposite to the $\p$-set method. The operation of the computer algorithm consists of minimizing the function
\begin{equation}
f(\{u_{nk}\},\alpha) = \| e^{i\alpha} U\{u_{nk}\}-U_g \|^2
\end{equation}
numerically in the space of all possible combinations of single-qubit gates $u_{nk}$, where the matrix norm is given by $\|A\|^2={\rm Tr}(A^\dagger A)$. The single-qubit gates $u_{nk}$ are parametrized by the three angles $\varphi_{nk}$, $\theta_{nk}$, and $\psi_{nk}$ according to the prescription
\begin{equation}
u_{nk} = \left(\begin{array}{c c}
\cos(\theta_{nk}) & -e^{i(\varphi_{nk}+\psi_{nk})}\sin(\theta_{nk}) \\
e^{i\varphi_{nk}}\sin(\theta_{nk}) &  e^{i\psi_{nk}} \cos(\theta_{nk})
\end{array}\right).
\end{equation}
Including the global phase $\alpha$ we count $3M(N+1)+1$ real parameters.
If the numerical search yields a minimum
\begin{equation}
f(\{u_{nk}\},\alpha) = 0,
\end{equation}
then the corresponding sequence Eq.~(\ref{minimizer_trial}) is identical to $U_g$.

The numerical results for $U_g=U_{\xor}$ and $X_n=S$ support the analytical result, i.e. there is no circuit for $N=1$. For $N=2$, we find a vast number of circuits other than Eq.~(\ref{cpf_circuit}) combined with Eq.~(\ref{basis_change}). For $U_g=E,E_T$, we do not find a solution for $N<4$, as guaranteed by the result of our previous analysis.

The impossibility of reducing the number of $S$ gates required for $E$ led us to the idea that it might be useful to replace $S$ by a three-qubit gate which is directly generated by the three-qubit Hamiltonian $H_3 = J({\bf S}_1\cdot {\bf S}_2 + {\bf S}_2\cdot {\bf S}_3 + {\bf S}_3\cdot {\bf S}_1)$, describing three simultaneously interacting spins with equal coupling constant $J$. We find that the analogue of $S$ for three spins is the gate
\begin{equation}
S_3 = e^{-i\pi/3}\exp(i \frac{4\pi}{9}  \frac{H_3}{J}),
\end{equation}
which is obtained when the interaction $J$ is switched on for time $\tau=4\hbar\pi/9J$. In analogy to Eq.~(\ref{s_operation}) we can express the action of $S_3$ on product states as
\begin{equation}
S_3 |\alpha_1 \alpha_2 \alpha_3 \rangle = -e^{i\pi/3}|\alpha_1 \alpha_2 \alpha_3 \rangle + \frac{e^{i\pi/6}}{2\sqrt{3}}\sum_{\sigma\in {\cal S}_3}|\alpha_{\sigma 1} \alpha_{\sigma 2} \alpha_{\sigma 3} \rangle ,
\end{equation}
where the second term is the symmetrization of the input state (${\cal S}_3$ denotes the permutation group of three objects). Whereas $S^4=\unity$, we find that $S_3^3 = -\unity$. In exact analogy to the `square-root of swap' gate, $\p(S_3)$ consists of states that have the form $|\alpha\rangle\otimes|\alpha\rangle\otimes |\alpha\rangle$, $|\alpha\rangle\in {\cal H}_2$. We can show that $E$ is not equal to any gate involving only one or two $S_3$. The case of three $S_3$ was studied numerically, but no circuit representation for $E$ was found.


\section{Parallel pulse mode}\label{parallel_pulse}

Operating a system described by the Hamiltonian $H(\vec{p})$ with parameters $\vec{p}=(J, {\bf B}_1, {\bf B}_2)$ given in Eq.~(\ref{hamiltonian}) as a quantum gate in the serial pulse mode is not optimal in the following sense: If several or all parameters $\vec{p}$ can be changed simultaneously, we expect that a given quantum gate, say $\xor$, can be performed faster than by changing only one parameter at a time as in the serial pulse mode. Generally, all parameters $\vec{p}$ are arbitrary functions of time such that the time evolution operator after time $t$ is a functional in $\vec{p}$ given by the time-ordered exponential
\begin{equation}
U_t[\vec{p}(\tau)] = T\exp\left(\frac{i}{\hbar}\int_0^t H(\vec{p}(\tau))\,d\tau\right).\label{propagator_general}
\end{equation}
Given some quantum gate $U_g$, we would now like to solve the integral equation $U_t[\vec{p}(\tau)]=U_g$ for the functions $\vec{p}(\tau)$. For unrestricted time $t$ and unbounded functions $\vec{p}(\tau)$ we immediately know how to construct such a solution by using the known universal set of gates Eqs.~(\ref{single_qubit}) and (\ref{sqrt_swap}) in the serial pulse mode. In general, this is not the optimal solution of $U_t[\vec{p}(\tau)]=U_g$. An optimal solution is given by a set of bounded functions $|p_i(\tau)|<M_i$ requiring minimal time $t$ for a fixed set of bounds $M_i$. Since it is not feasible to find an optimal solution among all such bounded functions, we will restrict ourselves to piecewise-constant functions. Splitting up the time interval $t$ into $N\geq 1$ parts, we write
\begin{eqnarray}
U_N(\vec{p}^{\,(1)},...,\vec{p}^{\,(N)}; \phi) &=& e^{i\phi} U_N(\vec{p}^{\,(N)})\cdots U_2(\vec{p}^{\,(2)}) U_1(\vec{p}^{\,(1)}),\nonumber\\
U_k(\vec{p}^{\,(k)})  &=& \exp\left\{2\pi i H(\vec{p}^{\,(k)})\right\}.\label{propagator_discrete}
\end{eqnarray}
For every time ``slice'', we have the freedom to choose a new set of parameters $\vec{p}^{\,(k)}=(J, {\bf B}_1, {\bf B}_2)$. Note that we allow for an arbitrary total phase $\phi$. By discretizing the problem in this way we have reduced the free parameters in the problem from the $P$ functions $p_i(t)$ to $P N$ real parameters $p_i^{(k)}$, $i=1,...,P$, $k=1,...,N$ where $P$ denotes the number of parameters $\vec{p}$ (in the case of the Heisenberg Hamiltonian Eq.~(\ref{hamiltonian}), $P=7$). The functions $p_i(t)$ are related to the discrete parameters $p_i^{(k)}$ through the relation
\begin{equation}\label{parameters}
p_i(t)=\frac{2\pi\hbar}{\tau_k}p_i^{(k)},\quad t_{k-1}\leq t<t_{k},
\end{equation}
where $\tau_k=t_k-t_{k-1}$ and $t_0=0$, $t_N=t$;
the time step $\tau_k$ has been absorbed into the dimensionless parameters $p_i^{(k)}$.
Once the problem is discretized, it becomes suitable for numerical treatment using the minimizer algorithm presented in Section \ref{serial_pulse}, minimizing the function $\|U_g-U_N\{p_i^{(k)}; \phi\}\|^2$ with respect to the $PN+1$ parameters $p_i^{(k)}$ and $\phi$. We try to find a solution to $U_N\{p_i^{(k)};\phi\}=U_g$ starting from $N=1$ and then increasing $N$ in unit steps. In practice, $N$ is limited by the available computational resources.

One approach to the problem would then be to fix $N$ and $t_i$
(e.g. use time steps of equal size, $\tau_k=\tau=t/N$). Then, the
constraint $|p_i(\tau)|<M_i$ implies $|p_i^{(k)}|<\tau_k
M_i/2\pi\hbar$.  In the following, however, we solve
$U_N\{p_i^{(k)};\phi\}=U_g$ with fixed $N$ (chosen as small as possible)
without any constraint for $p_i^{(k)}$ and then calculate $t$ for
given bounds $M_i$ using the formula (cf. Eq.~(\ref{parameters}))
\begin{equation}
t=\sum_{k=1}^N \tau_k = \sum_{k=1}^N \max_i\left(\frac{2\pi\hbar}{M_i}\left|p_i^{(k)}\right|\right).
\label{time}
\end{equation}
The parameter $p_i^{(k)}$ with the largest $p_i^{(k)}/M_i$ ratio determines the switching time for the $k^{\rm th}$ step.

\subsection{$\xor$ gate}
We now want to use this method for finding a pulse sequence $\vec{p}^{\,(k)}$ that generates the quantum $\xor$ gate, Eq.~(\ref{XOR}). Since $\xor$ is the same as the conditional phase flip ($\cpf$) up to the basis change Eq.~(\ref{basis_change}), we will first try to generate $\cpf$.
In the $S_z$ basis, the Heisenberg Hamiltonian $H(J,{\bf B}_1,{\bf B}_2)$, Eq.~(\ref{hamiltonian}), can be written in the following matrix form:
\begin{equation}
\frac{1}{2}\left(\begin{array}{c c c c}
B_1^z+B_2^z   & B_2^x-i B_2^y  & B_1^x-iB_1^y   & 0\\
B_2^x+i B_2^y & B_1^z-\!B_2^z-\!J & J              & B_1^x-i B_1^y\\
B_1^x+i B_1^y & J              & B_2^z-\!B_1^z-\!J & B_2^x-i B_2^y\\
0             & B_1^x+i B_1^y  & B_2^x+i B_2^y  & -B_1^z-B_2^z
\end{array}\right),\label{hamiltonian_matrix}
\end{equation}
where an irrelevant constant energy contribution is omitted.
We find analytically that $\cpf$ can be obtained in one time-step ($N=1$), i.e. for constant parameters $\vec{p}$,
\begin{eqnarray}
U_{\sc CPF} &=& \exp\left[2 \pi i H(J,{\bf B}_1,{\bf B}_2)\right],\nonumber\\
J &=& k-n-2m-\frac{1}{2},\nonumber\\
{\bf B}_1 &=& \frac{1}{2}(0,0,n+\frac{1}{2}+\sqrt{k^2-J^2}),\label{CPF_solution}\\
{\bf B}_2 &=& \frac{1}{2}(0,0,n+\frac{1}{2}-\sqrt{k^2-J^2}),\nonumber\\
\phi &=& -\pi (n+\frac{1}{2})\nonumber
\end{eqnarray}
where $n$, and $m$ are arbitrary integers, and $k$ is an integer satisfying $2|k|\ge |n+2m+\frac{1}{2}|$.
This solution is obtained by setting $B_i^x=B_i^z=0$, and diagonalizing
the resulting Hamiltonian matrix
\begin{equation}\label{CPF_matrix}
H = \frac{1}{2}\left(\begin{array}{c c c c}
A & 0    & 0    & 0\\
0 & -J+B & J    & 0\\
0 & J    & -J-B & 0\\
0 & 0    & 0    & -A\end{array}\right),
\end{equation}
where $A=B^z_1+B^z_2$ and $B=B^z_1-B^z_2$. The conditional phase flip $\cpf$ is invariant under the basis change that diagonalizes Eq.~(\ref{CPF_matrix}), and we can then solve the equation $\exp(i\phi)\exp(2\pi i H)=U_{\cpf}$ in the basis where both $H$ and $U_{\cpf}$ are diagonal. This yields the four equations $e^{\pi i A}=e^{2\pi i\lambda_+}=e^{2\pi i\lambda_-}=-e^{-\pi i A}=e^{-i\phi}$, where $\lambda_\pm=(-J\pm\sqrt{J^2+B^2})/2$ and $\pm A/2$ are the eigenvalues of $H$.
From these four equations we obtain the result Eq.~(\ref{CPF_solution}) for $\phi$, $A=B^z_1+B^z_2$, $B=B^z_1-B^z_2$, and $J$. Applying the basis change $V$ from Eq.~(\ref{basis_change}) to these solutions, we can build $\xor$ with a total of $N=3$ steps. The integers $k$, $m$, and $n$ can be chosen such that the switching time Eq.~(\ref{time}) is minimal for a given set of constraints $M_J$, $M_{B_1}$, $M_{B_2}$. In the specific case where all constraints are equal to $M$, we find that the solution for $k=1$, $m=n=0$,
\begin{equation}\label{cpf_parallel}
J=\frac{1}{2},\quad B_1^z=\frac{1}{4}(1+\sqrt{3}),\quad B_2^z=\frac{1}{4}(1-\sqrt{3}),
\end{equation}
has the shortest switching time,
\begin{equation}
t_{\cpf,p}=\frac{2\pi\hbar}{4 M}(1+\sqrt{3})=0.683 \frac{2\pi\hbar}{M},
\end{equation}
less than half the time which is used for the serial pulse quantum circuit Eq.~(\ref{cpf_circuit}), $t_{\cpf,s}=1.5\cdot 2\pi\hbar/M$.
Note that since the coupling is isotropic, the same gate (in a rotated basis) can be achieved with a magnetic field along any desired direction.
In order to obtain the $\xor$ gate, we must spend in addition twice the time $0.25\cdot 2\pi\hbar/M$ for the basis change $V$, Eq.~(\ref{basis_change}). The total switching time is then
\begin{equation}
t_{\xor,p}=\frac{2\pi\hbar}{4 M}(3+\sqrt{3})= 1.183\frac{2\pi\hbar}{M},
\end{equation}
about $59\%$ of the time required for the serial pulse quantum circuit, Eq.~(\ref{cpf_circuit}), including the change of basis Eq.~(\ref{basis_change}), $t_{\xor,s}= 2\cdot 2\pi\hbar/M$.
Of course, the basis change applied here is again a `serial pulse' action and therefore not optimal. We therefore study Eq.~(\ref{propagator_discrete}) directly for the $\xor$ gate, without using the $\cpf$ gate. It turns out that no solution exists for $N=1$. For $N=2$ our optimizer algorithm finds the numerical solution
\begin{equation}
U_{\xor}=e^{i\phi}\exp\left[2\pi i H(\vec{p}^{\,(2)})\right]\exp\left[2\pi i H(\vec{p}^{\,(1)})\right],
\end{equation}
with the parameter values\cite{footnote01}
\begin{equation}\label{xor_parallel}
\begin{array}{| r | r r r r r r r |}
\hline
k & J^{(k)} & B_{1x}^{(k)} & B_{2x}^{(k)} & B_{1y}^{(k)} & B_{2y}^{(k)} & B_{1z}^{(k)} & B_{2z}^{(k)}\\ \hline
1 &     0.187       &  -0.025 &       {\bf 0.464} &  0.205   & 0.195        & -0.420      & 0.395\\ 
2 &     {\bf 0.617} &  -0.220 &       0.345 & -0.384         & 0.244        & 0.353       & 0.108\\
\hline
\end{array}
\end{equation}
and the global phase $\phi=-0.8481 \cdot \pi$.
The total switching time for equal bounds is in this case $t_{\xor,p}=(0.4643+0.6170)2\pi\hbar/M=1.0813\cdot 2\pi\hbar/M$, compared to $t_{\xor,s}=2\cdot 2\pi\hbar/M$ for the serial switching. The numbers in boldface in Eq.~(\ref{xor_parallel}) indicate which parameter limits the switching time in each step. The solution Eq.~(\ref{xor_parallel}) appears to be a unique optimum for the case $N=2$.

\subsection{Three bit encoder $E$}
We can further parallelize the three-bit encoder $E$, Eq.~(\ref{e_circuit}). Instead of concatenating two XOR gates (which may or may not be produced using parallel pulses) we now try to find a more efficient parallel pulse sequence for $E$, given a system of three qubits which exhibit pairwise couplings among each other that can all be switched on simultaneously. The Hamiltonian for this three-spin system can be written as
\begin{equation}
H= \sum_{1\le i < j\le 3}J_{ij}{\bf S}_i\cdot {\bf S}_j + \sum_{i=1}^3{\bf B}_i\cdot {\bf S}_i.
\end{equation}
We find that there is a representation of the three-bit error correction encoder $E$ which consists of three parallel pulses only, instead of the four which it takes to perform two sequential XOR gates.
The following parallel pulse sequence produces $E$ up to a global phase $\phi=\pi/2$:
\begin{equation}
\begin{array}{| r | r r r | r r r |}
\hline
k & J_{12}^{(k)}  & J_{23}^{(k)}& J_{13}^{(k)} & B_{1x}^{(k)} & B_{2x}^{(k)} & B_{3x}^{(k)}\\ \hline
1 & 0.0000 & {\bf 8.2500}  & 0.0000 & 1.1153 &      6.1737  &      6.1739 \\
2 &    -0.9256 &      -5.3608 &      0.7863  & {\bf 5.7603} &   5.2422   &     1.7475 \\
3 &    0.0000  &    -1.7500    &   0.0000   &     0.4345 &     {\bf 3.5255}   &     {\bf 3.5255} \\ \hline
\hline
k & B_{1y}^{(k)} & B_{2y}^{(k)} & B_{3y}^{(k)}& B_{1z}^{(k)} & B_{2z}^{(k)}& B_{3z}^{(k)}\\ \hline
1   &  -1.6737 &   -0.2263 &      0.2262   &     1.1153   &     1.4649    &   -1.4649\\    
2 &            0.0000 &        0.0000   &    0.0000  &     0.0000    &    0.0000     &  0.0000\\
3  &     2.1709 & 1.4118  &      1.4118     &   0.4345   &     1.2560   &    1.2560 \\ \hline
\end{array}\label{e_parallel}
\end{equation}
For equal bounds the total switching time of $t_{E,p}=17.54\cdot 2\pi\hbar/M$ is much larger than the 4-pulse time $2t_{\xor,p}=2.163\cdot 2\pi\hbar/M$. Note that a better 3-pulse solution was not found, but cannot be excluded.


\section{Anisotropic systems}\label{anisotropic}

Systems where the spin-spin coupling is anisotropic are not described by the Heisenberg Hamiltonian Eq.~(\ref{hamiltonian}) that we studied as a generator for quantum gates in the previous sections.
In the two most notable cases, the Ising and the XY systems, it is known that universal quantum computation is possible. In the case of a system described by the Ising Hamiltonian $H_I = J S_1^z S_2^z$ and a homogeneous magnetic field in $z$ direction, there is a particularly simple realization of the $\cpf$ gate with constant parameters, namely $U_{\cpf} = \exp(i\pi(1-2S_1^z-2S_2^z+4S_1^z S_2^z)/4)$ \cite{loss}. One might be tempted by this to `transform' the Heisenberg interaction Eq.~(\ref{hamiltonian}) into an Ising interaction by adding time-dependent fields $H_0(t)={\bf B}_1(t)\cdot {\bf S}_1 + {\bf B}_2(t)\cdot {\bf S}_2$ to the coupling Hamiltonian $V(t)=J(t){\bf S}_1\cdot {\bf S}_2$ such that the coupling in the interaction picture, $V_I(t)=U(t)V(t)U(t)^\dagger$, with $U(t)=T\exp(i\int_0^t H_0(\tau)d\tau)$, would be identical to the Ising coupling $V_I(t)=H_I$, or even to switch the coupling off and on using this method, i.e. $V_I(t)=0$ for a certain choice of $H_0(t)$. It turns out that this is impossible, since the `transformed' coupling must have the form
\begin{equation}\label{trsf_coupling}
V_I(t) = J(t) {\bf S}_1\cdot (R(t) {\bf S}_2),
\end{equation}
where $R(t)$ is a time-dependent rotation matrix.
For $J\neq 0$, this clearly excludes the complete `switching off' of the interaction.
Furthermore, the coupling Eq.~(\ref{trsf_coupling}) is still isotropic at every instant $t$.

In spite of the impossibility of using a Heisenberg system as an effective XY (or Ising) system by adding time-dependent fields, there are XY systems in nature which have been proposed for quantum computation\cite{imamoglu}. In the case of Ising systems we have seen above that there is a very simple prescription for generating the $\xor$ gate. We devote the rest of this Section to demonstrating that $\xor$ can also be obtained with XY coupling.
For two spins with $s=1/2$ the XY Hamiltonian is given by
\begin{equation}\label{xy_hamiltonian}
H_{XY} = J (S_1^x S_2^x + S_1^y S_2^y)
= \frac{J}{2} \left(\begin{array}{c c c c}
0 & 0 & 0 & 0\\
0 & 0 & 1 & 0\\
0 & 1 & 0 & 0\\
0 & 0 & 0 & 0
\end{array}\right),
\end{equation}
where for the matrix representation we chose the $S^z$ basis.
The corresponding time evolution operator is
\begin{equation}
U_{XY}(\phi=Jt) = \exp(itH_{XY})
= \left(\begin{array}{c c c}
1 & & \\
  & e^{i\phi S^x} & \\
  & & 1
\end{array}\right).
\end{equation}
There is a qualitative difference between two qubits coupled via an XY and ones coupled by a Heisenberg interaction: it is impossible to generate powers $U_{\rm swap}^{\alpha}$ ($0<\alpha<1$) of the swap gate Eq.~(\ref{swap}) with only one use of $U_{XY}(\phi)$ together with single-qubit operations. In particular, this is impossible for the `square-root of swap' gate $U_{\rm swap}^{1/2}$. In spite of this, we found that the $\cpf$ gate can be produced by the serial-pulse sequence
\begin{eqnarray}
U_{CPF}&=&e^{i\pi/4}e^{2 i\pi {\bf n}_1\cdot{\bf S}_1/3}
e^{2 i\pi {\bf n}_2\cdot{\bf S}_2/3}\nonumber\\ 
&\times& U_{XY}(\pi/2) e^{i\pi S_1^y}U_{XY}(\pi/2)
e^{-i\pi S_1^x/2}e^{-i\pi S_2^x/2},\label{cpf_circuit_xy}
\end{eqnarray}
where ${\bf n}_1=(1,-1,1)/\sqrt{3}$ and ${\bf n}_2=(1,1,-1)/\sqrt{3}$. This is a proof that XY systems with single-qubit interactions are in principle capable of universal quantum computation.

We now consider parallel switching with the XY dynamics,
\begin{eqnarray}
H_{XY,B}    &=& H_{XY}+{\bf B}_1\cdot {\bf S}_1+{\bf B}_2\cdot {\bf S}_2,\\
U_{XY,B}(t) &=& \exp(itH_{XY,B}).
\end{eqnarray}
As in the case of Heisenberg interactions, we first consider the $\cpf$ gate
which can be used to assemble the $\xor$ gate as shown in Eq.~(\ref{basis_change}).
We have not found a possibility to generate the $\cpf$ gate
Eq.~(\ref{CPF}) with the XY Hamiltonian with  applied magnetic fields with
constant parameters ($N=1$) using a numerical search\cite{footnote1}.
If the switching is performed in two steps ($N=2$), we  find numerically
that there are several possibilities to generate $U_{CPF}$ in the form
\begin{equation}
U_{CPF} = e^{i\phi}\,U_2 U_1,
\end{equation}
where $U_k=\exp\left[2\pi i H_{XY,B}(J^{(k)},B_x^{(k)},B_z^{(k)})\right]$,
$k=1,2$. Note that all magnetic fields can be chosen homogeneous (${\bf
B}_1^{(k)}={\bf B}_2^{(k)}\equiv{\bf B}^{(k)}$) and perpendicular to the
$y$-axis ($B_y=0$). 
Here we give one possible realization which is found numerically ($\phi=-3\pi/4$):
\begin{equation}\label{cpf_parallel_xy}
\begin{array}{| r | r r r |}
\hline
k &  J^{(k)}  &  B_x^{(k)} &  B_z^{(k)}\\ \hline
1 &  0.7500 &  {\bf 0.7906}  &  0.5728\\
2 &  {\bf 0.5000} &  0.0000  &  0.2500\\ \hline
\end{array}
\end{equation}
The total switching time for $\cpf$, assuming equal bounds $M_J=M_B\equiv M$ for $J$ and $B$, is $t_{\cpf,p}^{XY}=1.291\cdot 2\pi\hbar/M$, compared to $t_{\cpf,s}^{XY}=2.167\cdot 2\pi\hbar/M$ for the serial pulse sequence defined in Eq.~(\ref{cpf_circuit_xy}).

In order to produce the XOR gate Eq.~(\ref{XOR})
we can implement the basis change Eq.~(\ref{basis_change})
using the single-qubit rotation $V$.
This procedure requires a total of four steps for the XOR gate.
Another way of achieving XOR is the following sequence which
we found numerically and which takes only three steps:
\begin{equation}
U_{XOR} = \exp(3i\pi/4)U_3  U_2  U_1,
\end{equation}
with the following parameters:

\begin{equation}\label{xor_parallel_xy}
\begin{array}{| r | r r r r r r r |}
\hline
  k &   J^{(k)}    &    B_{1x}^{(k)} & B_{2x}^{(k)} &   B_{1y}^{(k)} & B_{2y}^{(k)} & B_{1z}^{(k)} & B_{2z}^{(k)}\\
\hline
  1 &   1.802   &    0.615     &  {\bf 2.045}    &   0.020     &  0.316    &   0.794   &    0.130 \\
  2 &   {\bf 3.344}   &    0.348     &  0.718    &   0.259     &  0.493    &   1.583   &    1.062 \\
  3 &   {\bf 1.903}   &    1.193     &  0.705    &   0.413     &  -0.305   &   0.589   &    0.604 \\\hline
\end{array}
\end{equation}
The total switching time of $t_{\xor,p}^{XY}=7.29\cdot 2\pi\hbar/M$ (compared to $2.67\cdot 2\pi\hbar/M$ using $\cpf$ and a basis change) indicates that Eq.~(\ref{xor_parallel_xy}) is not an optimal solution.

\section{Requirements for Parallel Switching}\label{requirements}
The parallel switching mechanism presented in the Sections \ref{parallel_pulse} and \ref{anisotropic} relies on the following
essential assumptions:

(a) Each of the parameters in the Hamiltonian can be varied independently.
That is, the coupling can be varied independent of the magnetic fields in the Hamiltonian Eq.~(\ref{hamiltonian}).

(b) We know the exact relation between the externally controlled parameters (such as the electric
and magnetic field or a gate voltage) and the parameters in the Hamiltonian.

(c) The switching is synchronous, with all parameters $p_i$ varying with the same time profile $p_i(t)=\tilde{p}_i f(t)$. The change of parameters does not have to be step-like, but can be chosen to have some smooth pulse form. Also, any pulse magnitudes $\tilde{p}_i$ are allowed.

Whether the above requirements can be fulfilled depends on the underlying microscopic mechanisms which are
responsible for the effective Hamiltonian, such as the Heisenberg Hamiltonian, Eq.~(\ref{hamiltonian}).
In our previous work\cite{burkard} we have used the model Hamiltonian
\begin{eqnarray}
H &=& \frac{1}{2m} \sum_{i=1,2} \left[\left({\bf p}_i-\frac{e}{c}{\bf A}({\bf r}_i)
\right)^2+ex_iE\right.\nonumber\\
&+&\left. \frac{m\omega_0^2}{2}\left(\frac{1}{4 a^2}\left(x_i^2-a^2
\right)^2+y_i^2\right)\right]+\frac{e^2}{\kappa\left| {\bf r}_1-{\bf r}_2\right|},\label{microscopic}
\end{eqnarray}
with ${\bf A}({\bf r})=B(-y,x,0)/2$ to describe the orbital dynamics of electrons in coupled
quantum dots.
Here, ${\bf r}_i$ and ${\bf p}_i$ denote the location and momentum of the electron $i$ which is moving
in two dimensions in a double-well potential $V$ with characteristic energy $\hbar\omega_0$ and a
magnetic field $B$ perpendicular to the 2D electron system and an electric field $E$ parallel to
the coupling axis of the two wells. The distance between the quantum dots is denoted by $2a$, the 
effective mass and the charge of the electron by $m$ and $e$, and the dielectric constant of the
material by $\kappa$. As we pointed out earlier\cite{burkard}, the spin-orbit interaction
$H_{\rm so}=(\omega_0^2/2m_0 c^2){\bf S}\cdot {\bf L}$ is very small for an electron in a
parabolically confined quantum dot. Note however that this expression for the spin-orbit
coupling contains the bare electron mass $m_0$ instead of the effective electron mass $m$,
making the spin-orbit coupling in GaAs smaller than estimated in \onlinecite{burkard}
by a factor of $m_0/m \simeq 15$\cite{footnote2}.
For a quantum dot with confining energy $\hbar\omega_0=3\,{\rm meV}$,
we obtain $H_{\rm so}/\hbar\omega_0\approx 10^{-8}$.

Concerning condition (a) we have found that the spin-spin coupling $J$ can be controlled by several
external ``knobs''.
The gate voltage $V$ applied between the coupled quantum dots controls the height of the
barrier for tunneling of an electron from one dot into the other and therefore strongly influences the
exchange coupling $J$ between the electronic spins. In a similar manner, $J$ depends on the inter-dot
distance $2a$. We have also found\cite{burkard} that an external magnetic field $B$
perpendicular to the 2DEG causes a
strong change (even a sign reversal) of $J$. Not surprisingly, an electric field $E$ applied along
the coupling direction of the dots also changes the exchange coupling, which can be understood as an effect of
level detuning. When switching on a magnetic field, the effect of the field on $J$ could be compensated by changing another independent control parameter,
e.g. the electric field. In practice, one has to know the functional dependence $J(V,a,B,E)$ in the range
where it is used, see also (b).

While a magnetic field perpendicular to the 2DEG strongly influences the exchange $J$, we can argue
that sufficiently weak in-plane magnetic fields have little influence on $J$.
Classically, the motion of a particle in a plane is not affected by a magnetic field in the plane,
since the Lorentz force is orthogonal to the plane. Quantum mechanically, we can describe a particle
in a magnetic field confined to a plane by the Hamiltonian
\begin{equation}\label{well_field}
H = \frac{1}{2m}\left({\bf p}-\frac{e}{c}{\bf A}\right)^2 + \frac{m\omega^2}{2}z^2,
\end{equation}
where the vector potential ${\bf A}=B(0,-z,0)$ corresponds to a magnetic field of magnitude $B$ along
the $x$ axis and the confinement in z direction is modeled by a harmonic potential with frequency $\omega$.
In this gauge, the Hamiltonian can be rewritten in the form
\begin{equation}
H = \frac{p_x^2+p_z^2}{2m} +\frac{p_y^2}{2\bar{m}} + \frac{m\bar{\omega}^2}{2} (z-z_0)^2,
\end{equation}
with the renormalized effective mass in y-direction, $\bar{m}=m(1+4\omega_L^2/\omega^2)$, the renormalized
confining energy $\hbar\bar{\omega}=\hbar\omega\sqrt{1+4\omega_L^2/\omega^2}$, and a shift in the
confining potential $z_0=2p_y\omega_L/m\bar{\omega}^2$ which depends on the momentum $p_y$ in the $y$
direction and the Larmor frequency $\omega_L=eB/2mc$. Note that the corrections due to the magnetic field
in the resulting 2D Hamiltonian
\begin{equation}
H_{2D} = \frac{p_x^2}{2m}+\frac{p_y^2}{2\bar{m}},
\end{equation}
are of the order $\omega_L^2/\omega^2$ or $(a_z/l_B)^4$, where $l_B=\sqrt{\hbar c/eB}$ denotes the
magnetic length and $a_z=\sqrt{\hbar/m\omega}$ the confinement length. Usually, we are interested in
the case of strong confinement and moderate magnetic fields where $a_z\ll l_B$, therefore
$\bar{m}\approx m$ up to small corrections. In this case, an in-plane magnetic field does
not affect the orbital degrees of freedom of the 2D electrons.

The condition (b) can be fulfilled in two ways. Either we have a theoretical description of the
dependence of the Hamiltonian parameters ($J$, ${\bf B}_i$) on the 
control parameters ($V$, $a$, $B$, $E$) or this relation is first mapped out in
experiment and the obtained data is used later for the control of the device. 
A good approximate description is possible in the case of adiabatic switching. In order 
to demonstrate this, we cast the microscopic
Hamiltonian into the form $H(t)=H_0+V(t)$. Then we find the instantaneous eigenstates $|n(t)\rangle$ and
the corresponding instantaneous eigenvalues $\epsilon_n(t)$ by solving the time-independent Schr\"odinger
equation for fixed time $t$.
The instantaneous eigenstate $|n(t)\rangle$ is a good approximation for the time evolution of the initial
state $|n(0)\rangle$, provided the adiabaticity criterion
\begin{equation}\label{adiabaticity}
\left|\frac{\langle m|\partial_t V|n\rangle}{\epsilon_m-\epsilon_n}\right|\ll \frac{1}{\tau_s},
\end{equation}
is met, where $\tau_s$ denotes the switching time. Eq.~(\ref{adiabaticity}) means that the change of the
external control parameters during the switching time should be much smaller than the level spacing in
the microscopic Hamiltonian. In the case of coupled quantum dots in the adiabatic regime,
$J(t)=\epsilon_t(t) - \epsilon_s (t)$
is the level spacing between the instantaneous singlet and triplet energies.

Note also that if $V(t)$ respects some symmetry, there can be selection rules
that make Eq.~(\ref{adiabaticity}) less stringent.
In the case of two coupled quantum dots with an applied homogeneous magnetic field the total spin is
conserved by $V(t)$ and therefore only transitions to higher orbital levels of the quantum dots are
relevant. Therefore, the less stringent condition
$1/\tau_s\approx |\dot{V}/V|\ll \Delta\bar{\epsilon}/\hbar$ is
sufficient for adiabatic switching\cite{burkard}. Here, $\Delta\bar{\epsilon}$ denotes the orbital level
distance averaged over the switching time. Since in this case the Zeeman energy is independent of the
space coordinates, it commutes with the orbital Hamiltonian and does not affect adiabaticity.
The case of inhomogeneous magnetic fields is more intricate.
The lack of a selection rule enforces the more stringent adiabaticity condition\cite{burkard}
$1/\tau_s\approx |\dot{V}/V|\ll \bar{J}/\hbar\ll \Delta\bar{\epsilon}/\hbar$, where $\bar{J}$ denotes
the average exchange coupling
during the switching. In addition to this, the Zeeman term does not commute with the orbital Hamiltonian
in the case of inhomogeneous fields and therefore also influences $J$. Due to these difficulties, we
presently do not know how to calculate the parameter $J$ in Eq.~(\ref{hamiltonian}) in the presence of an
inhomogeneous field, ${\bf B}_1\neq {\bf B}_2$.

The condition of synchronous switching (c) is mainly a technical issue.
We would like to stress that the choice of the pulse form has a decisive influence on 
whether the adiabaticity condition Eq.~(\ref{adiabaticity}) can be satisfied or not.
It is quite easy to see that a rectangular pulse is unsuitable because it has
infinite derivatives. Both Gaussian ($\exp(-t^2/\Delta t^2)$) and exponential
($\exp(-|t|/\Delta t)$) pulses are far better than a rectangular
pulse. The exponential pulse has the advantage that $|\dot{V}/V|$ is independent
of $t$ compared to the Gaussian pulse where $|\dot{V}/V|\propto t$.
However, the exponential pulse has the disadvantage that it has a cusp at $t=0$
which causes algebraically decaying tails in its Fourier spectrum.
We can combine the advantages of both pulses by using the sech pulse,
${\rm sech}(t/\Delta t)=1/{\rm cosh}(t/\Delta t)$.
Since all the pulses have to be cut off at some finite time $\pm\tau_s/2$, we choose
the width of the pulse $\Delta t$ smaller than the actual switching time $\tau_s$,
i.e. we choose $\alpha=\tau_s/\Delta t>1$.
By substituting the sech pulse into the adiabaticity condition
Eq.~(\ref{adiabaticity}), we obtain the condition $\tau_s\gg\alpha \hbar/\Delta\bar{\epsilon}$
in the case where the spin is conserved (homogeneous magnetic field) and
$\tau_s\gg\alpha \hbar/\bar{J}$ otherwise.

\section{Applications}\label{applications}
We will now give a detailed description of how a system of three
coupled quantum dots could be controlled in order
to test the functionality of three bit quantum error correction in 
that system. We denote the maximal
coupling and magnetic field that can be applied by $J_{\max}$ and $B_{\max}$.
If only one of the parameters $J_{ij}$, ${\bf B}_{i}$ can be made non-zero at a 
given instant, then the following serial-pulse sequence has to be applied:
\begin{equation}\label{instructions_serial}
\begin{array}{| r | c c r || r | c c r |}
\hline
{\rm step} & {\rm duration} & {\rm parameter} & {\rm value}\\ \hline
1  & \tau_{B}/4 & B_y^2 & B_{\max} \\ 
2  & \tau_{J}/4 & J_{12} & J_{\max}\\ 
3  & \tau_{B}/2 & B_z^1 & B_{\max} \\ 
4  & \tau_{J}/4 & J_{12} & J_{\max}\\ 
5  & \tau_{B}/4 & B_z^1 & B_{\max} \\ 
6  & \tau_{B}/4 & B_z^2 & -B_{\max}\\ 
7  & \tau_{B}/4 & B_y^2 & -B_{\max}\\ 
8  & \tau_{B}/4 & B_y^3 & B_{\max}\\
9  & \tau_{J}/4 & J_{13} & J_{\max}\\
10 & \tau_{B}/2 & B_z^1 & B_{\max}\\
11 & \tau_{J}/4 & J_{13} & J_{\max}\\
12 & \tau_{B}/4 & B_z^1 & B_{\max}\\
13 & \tau_{B}/4 & B_z^3 & -B_{\max}\\
14 & \tau_{B}/4 & B_y^3 & -B_{\max}\\ \hline
15 & \tau_{n} & B_x & {\rm random}\\ \hline
16-29 & {\rm repeat} & {\rm 1-14} & \\ \hline
\end{array}
\end{equation}
where $\tau_{J}=2\pi\hbar/J_{\max}$ and 
$\tau_{B}=2\pi\hbar/g\mu_{B}B_{\max}$. Step 15 describes the artificial
introduction of noise into the system by applying a random
magnetic field in the $x$ direction, causing random spin flips in a
time $\tau_n\lesssim \pi/g\mu_B\bar{B}_x$, where $\bar{B}_x$ denotes the 
mean amplitude of the random $B$ field. 
After step 29 is completed, qubits 2 and 3 are measured and
qubit 1 is flipped (by applying $B_x^1=B_{\max}$ for time $\tau_{B}/2$) if
both measurements yield 1 (spin down).
The total switching time for steps 1 to 29 
then amounts to $\tau_s = 6\tau_B + 2\tau_J + \tau_n$.

In a device where parallel pulses are possible, i.e.\ where the conditions (a)-(c)
from Sec.~\ref{requirements} are fulfilled, the following pulse sequence can be
applied with the same effect:
\begin{equation}\label{instructions_parallel}
\begin{array}{| r | r | r r r r r r r |}
\hline
i & \tau_i & & & & & & & \\ \hline
    &               & J_{12} & B_{1x} & B_{2x} & B_{1y} & B_{2y} & B_{1z} & B_{2z}\\ \hline
1   & .464 \,\tau &   .402 & -.054 & 1      &  .442      & .419        & -.905      & .851\\ 
2   & .617 \,\tau &   1      & -.356 & .559 & -.622      & .396        & .572       & .176\\ \hline
    &               &   J_{13} &         B_{1x} & B_{3x} & B_{1y} & B_{3y} & B_{1z} & B_{3z}\\ \hline
3   & .464 \,\tau &   .402 & -.054 & 1 &  .442   & .419        & -.905      & .851\\ 
4   & .617 \,\tau &   1      & -.356 & .559 & -.622      & .396        & .572       & .176\\ \hline
    &               &          & B_{1x} & B_{2x}  & B_{3x} & & & \\ \hline
5   & \tau_n        &          & {\rm rnd} &   {\rm rnd} &   {\rm rnd} & & & \\ \hline
6   &               &            & 1 & & & & & \\
|   & 2.162 \,\tau  &{\rm repeat}& | & & & & &\\
9   &               &            & 4 & & & & &  \\  \hline
\end{array}
\end{equation}
We have assumed that the maximal Zeeman energy is equal to the maximal coupling,
$g\mu_B B_{\max}=J_{\max}\equiv M$, and defined $\tau\equiv\tau_B=\tau_J$. All
parameters are given in units of $M$. The parameters in every step can be
multiplied by any pulse shape $f(t)$ with $\int_0^{\tau_i}f(t)dt=1$, where
$\tau_i$ denotes the duration of step $i$. Note that in every step, the pulse
shape has to be the same for all parameters.
Parameters that are omitted in Eq.~(\ref{instructions_parallel}) are set to zero.
The total switching time in this parallelized version amounts to $\tau_p=4.3252\,\tau + \tau_n$,
compared to $\tau_s=8\tau+\tau_n$ in the case of serial switching.

\section{Conclusion}\label{conclusion}

We have studied the minimal requirements for the implementation of the $\xor$ gate,
the conditional phase flip ($\cpf$) gate,
the encoding circuit $E$ used for three bit error correction, and the teleportation encoder $E_T$,
all for Heisenberg-coupled spins with $s=1/2$. In addition to this, we have also considered anisotropic
spin-spin coupling as described in the XY model. Two different methods for
generating quantum gates with a time-dependent Hamiltonian have
been discussed and compared, the ``conventional'' serial pulse method and a new
method involving parallel pulses.

The main results of our work are the parallel pulses for the conditional phase flip
(Eq.~(\ref{cpf_parallel})) and $\xor$ (Eq.~(\ref{xor_parallel})) using Heisenberg 
dynamics, and the corresponding results (Eq.~(\ref{cpf_parallel_xy}) and
Eq.~(\ref{xor_parallel_xy})) for XY dynamics. The direct parallel-pulse sequence
Eq.~(\ref{e_parallel}) for the three-bit encoder $E$ was found; however,
it is possible that a faster pulse sequence for this gate can be found with
more numerical effort.

The following results for serial switching have been found:
There is an analogue of the known circuit
Eq.~(\ref{cpf_circuit}) for $\cpf$ (cf.\ Fig.~\ref{cpf_fig}) for
systems with XY coupling, which is given in Eq.~(\ref{cpf_circuit_xy}).
For Heisenberg coupling, we have proved that the known
circuit Eq.~(\ref{cpf_circuit}) is optimal in the sense that
$\cpf$ cannot be obtained with one `square-root-of-swap' gate.
For the proof we invoked the set $\p(U)$ of all product states
that are mapped onto product states by a quantum gate $U$;
$\p(U)$ helps to distinguish quantum gates modulo concatenation of
single-qubit gates. The same tool was also used to prove that
the encoder $E$ for quantum error correction cannot be 
generated with serial switching with fewer than four
`square-root-of-swap' gates. The same is true for the
encoder $E_T$ for the teleportation of one qubit.

The results for the parallel-pulse $\xor$ for 
isotropic Heisenberg interactions and the results
for $\cpf$ and $\xor$ for XY interactions, Eqs.~(\ref{xor_parallel}),
(\ref{cpf_parallel_xy}), and (\ref{xor_parallel_xy}),
and for the three bit encoder Eq.~(\ref{e_parallel})
were all found using the computer algorithm described
in Section \ref{parallel_pulse}. This algorithm searches
for a (parallel) pulse sequence for an arbitrary 
quantum gate operating on any number of qubits. The
number of qubits and the complexity of the pulse
sequence that can be studied are only limited by the
available computational resources.

Quantum computations are very often presented in the form of
quantum circuits, i.e.\ as a sequence of gates belonging to a small set of universal gates.
Our examples of parallel-pulse gates illustrate that such quantum circuits are in general not the
most efficient way of performing a quantum computation. The reason for this is that standard quantum
circuits only allow the use of a small fraction of the possible time evolutions that can
be generated by the underlying Hamiltonian. While for the two- and three-qubit gates we have
studied here, we could optimize the switching time by typically a factor of about two by
using parallel pulses, 
it can be speculated that for gates operating on many qubits or whole quantum computations,
switching times could be  reduced by a much larger amount.
Note also that the parallel pulses we have studied here represent only a small
subset of the possible time evolutions themselves, since we have been restricted to very simple
discretized pulses of up to three time-steps.

While quantum circuits are very intuitive and provide an excellent framework for the theoretical
study of quantum algorithms and their connection to classical algorithms, the representation
of quantum gates or whole computations as parallel pulse sequences may turn out to be more efficient
for a number of physical implementations.

\acknowledgments
We would like to thank C.H. Bennett, A. Imamo\=glu, B.M. Terhal, and A.V. Thapliyal for
interesting discussions.
This work is supported by the Swiss National Science Foundation.
D.P. DiVincenzo acknowledges the funding under grant ARO DAAG55-98-C-0041.


\begin{thebibliography}{99}
\bibitem{footnote0}
For $M$ qubits, the set of product state is defined as
$\p = \left\{|\Psi\rangle\in{\cal H}={\cal H}_2^{\otimes M}\Big|
|\Psi\rangle = |\phi_1\rangle\otimes\cdots\otimes|\phi_M\rangle ;|\phi_i\rangle\in{\cal H}_2\right\}$.
\bibitem{footnote01}
For compactness, we have truncated our numerical results after the third or fourth decimal.
If desired, much higher precision can be obtained with our numerical algorithm.
\bibitem{footnote1}
In the special case where $B_{ix}=B_{iy}=0$
this can be demonstrated by applying the same procedure as for the Heisenberg
interaction, showing that the corresponding four equations have no solution.
\end{thebibliography}
\end{document}